\documentclass[aps,prd,showpacs,superscriptaddress,twocolumn,raggedfooter,raggedbottom]{revtex4-1}   

\usepackage{amssymb}
\usepackage{amsmath}
\usepackage{amssymb}
\usepackage{graphicx}
\usepackage{subfigure}
\usepackage{color}
\usepackage{mathrsfs} 
\usepackage{float}

\usepackage{soul}

\newcommand{\sech}{\operatorname{sech}}

\usepackage{hyperref}

\begin{document}

\title{On the Ground State Quantum Droplet for Large Chemical Potentials}

\author{J. Holmer}
\affiliation{Department of Mathematics, Brown 
University, Providence RI, 02912, USA}

\author{K.~Z. Zhang}
\affiliation{Department of Mathematics,
Northeastern University, Boston, MA, 02115, USA}

\author{P.~G.\ Kevrekidis}
\affiliation{Department of Mathematics and Statistics, University of Massachusetts, Amherst, MA 01003-4515, USA}

\begin{abstract}
In the present work we revisit the problem
of the quantum droplet in atomic Bose-Einstein
condensates with an eye towards describing
its ground state in the large density, so-called
Thomas-Fermi limit. We consider the problem
as being separable into 3 distinct regions:
an inner one, where the Thomas-Fermi approximation
is valid, a sharp transition region where the
density abruptly drops towards the (vanishing)
background value and an outer region which
asymptotes to the background value. 
We analyze the spatial extent of each of these
regions, and develop a systematic effective
description of the rapid intermediate transition
region. Accordingly, we derive a uniformly
valid description of the ground state that
is found to very accurately match our
numerical computations. As an additional application
of our considerations, we show that this formulation
allows for an analytical approximation of excited
states such as the (trapped) dark soliton in the large density
limit.
\end{abstract}

\maketitle

\section{Introduction}

Over the past few years, an emerging topic
in the physics of atomic Bose-Einstein 
condensates has been the exploration of
quantum droplets. The latter 
constitute self-bound states arising
from the interplay between the mean-field
and quantum fluctuation energetic 
contributions~\cite{petrov2015quantum}.
In this competition, 
it is also important to recognize
the role of the system's dimensionality~\cite{khan2022quantum}.
Early experimental observations of the relevant
settings took 
place in dipolar condensates~\cite{schmitt2016self,chomaz2022dipolar},
followed shortly thereafter by trapped bosonic
mixtures with contact interactions~\cite{cabrera2018quantum,cheiney2018bright,d2019observation}.
Such droplet states were also observed
in free space, e.g., in the work 
of~\cite{semeghini2018self}. 

In droplet-bearing systems,
the critical role of quantum fluctuations
has been theoretically incorporated by means
of the well-known Lee--Huang--Yang (LHY) correction term~\cite{lee1957eigenvalues} that is 
suitably added to the standard mean-field
cubic nonlinearity description~\cite{petrov2015quantum,Petrov_2016}.
This leads to an extended Gross-Pitaevskii
equation (EGPE) description that has been found
to be fruitful for the theoretical and
computational identification of such droplet
patterns~\cite{Ferioli2020dynamical,app11020866,mistakidis2022cold}. Accordingly, this 
formulation has been used in order
to describe modulational instability and
related features~\cite{mithun2019inter,mithun2021statistical,otajonov2022modulational}, 
collective excitations~\cite{PhysRevA.101.051601,englezos2023correlated,sturmer2021breathing,cappellaro2018collective} and nonlinear
wave structures in the form of solitary waves
and vortices~\cite{yongyao,gangwar2022dynamics,kartashov2022spinor,PhysRevResearch.5.023175,saqlain2023dragging,katsimiga2023solitary,condmat8030067}. Indeed, 
already this new form of ``liquid matter''
has been the subject of not only numerous
studies, but also relevant reviews
such as~\cite{luo2020new}.

In the case of the standard single-component
atomic condensates (with cubic nonlinearity), 
the study of the system's ground state
is quite mature at this stage. For the standard one-dimensional
GPE, a well-known approximation is that of the Thomas-Fermi (TF)
limit, which is progressively more accurate as the chemical
potential is increased and consists of an inverted
parabola (with compact support)~\cite{pethick2008bose,stringari}.
What is perhaps somewhat less well-known
in the physics community is that while this (TF)
description is accurate close to the center
of the trapped 1d condensate, there have
been some significant mathematical works
that have offered refinements in the vicinity
of the condensate edges. In the latter,
the dispersion becomes significant, creating
a boundary layer which requires a more
refined multiple-scales analysis to be
properly captured, as has been explored
in works such as those of~\cite{Gallo2009OnTT,karali}. Accordingly,
these works have been able to
 accurately approximate the relevant layer,
 by leveraging the so-called
Hastings–McLeod solution of the Painlev{\'e}-II equation. While there exist more accurate,
quasi-1d approximations of the full 3d
problem~\cite{salasnich,delgado}, this analysis is significant for
various purposes, including towards
understanding not only the asymptotics but also
the excitation spectrum
of the relevant ground state in an analytical
(or semi-analytical/asymptotic) form.

 At the present time, to the best of 
 our knowledge, there exists no analogous
 analysis of the problem of the quantum
 droplet. In the latter case, indeed, as
 we will see below, the issue of the 
 asymptotic state is further exacerbated
 by the attractive nature of the nonlinearity
 for small densities. In that light, the
 TF approximation fails already at a finite
 density and it is not possible to construct
 a corresponding simple-minded approximation
 profile by neglecting the wavefunction curvature.
 Accordingly, it is of interest to develop
 ideally a uniform approximation that allows
 us to capture the large chemical potential
 droplet wavefunction, in a way analogous
 to~\cite{Gallo2009OnTT,karali}. This is the aim
 of the present work. In order to do so, we 
 leverage a multiple-scales spatial analysis
 of the stationary state problem, separating 
 the region close to the center (where the TF
approximation turns out to still be valid),
an intermediate, steep-descent region that
we suitably quantify and finally the asymptotic
decay towards the background value. Combining these
3 regions, we eventually obtain a uniform 
expression for the quantum droplet profile
in the presence of a trap in the large chemical
potential limit. This expression is of value well-beyond
the strict confines of the ground state: indeed, we show
that it provides us with the tools to accurately approximate
excited states in the form, e.g., of the prototypical 
dark soliton state of the trapped system.

Our presentation will be structured as follows.
In section II we present the mathematical setup
of the problem and our main results for the droplet
system's ground state. 
In section III, we provide the
details of our multiple-scales analysis, while
in section IV we present an extension of the 
method for the case of the dark soliton 
configuration. 
Finally, in section V, we summarize our findings and
present our conclusions, as well as some
directions for future work.

\section{Mathematical Setup \& Main Result}

The framework of interest to us herein
will concern the homonuclear mass-balanced
case of a one-dimensional (1d) bosonic mixture of two
different (hyperfine) states.
In line with earlier works 
including~\cite{PhysRevResearch.5.023175,katsimiga2023solitary,condmat8030067}, we will assume that the two species
feature equal self-repulsion $g_{11}=g_{22}\equiv g$, while across the species the attraction
renders $g_{12}$ negative. 
A prototypical example thereof can be encountered
in the context, e.g., of $^{39}$K, whose
states $|1,-1\rangle$ and $|1,0\rangle$
have been considered previously~\cite{semeghini2018self} (in the
3d realm). Considering the relevant EGPE
model 

\begin{equation}
i u_{t}=-\frac{1}{2}u_{xx}+|u|^{2} u - \delta |u| u + V(x) u.
\label{jzeq1}
\end{equation}
Here, we have already assumed that 
the energy of the system is
measured in units of $\hbar ^{2}/(m\xi ^{2})$, 
and $\xi =\pi \hbar ^{2}\sqrt{|\delta g|}/(mg\sqrt{2g})$ is the healing length,
while the quantity $\delta g=g_{12}+g$
combines the inter- and intra-species 
scattering lengths. Additionally, 
in this formulation, time, length and
wave function are scaled according to $t'=\hbar/\left(m\xi ^{2}\right) $, $x'=\xi x$ and
$u'=(2\sqrt{g})^{3/2} u/(\pi \xi (2|\delta g|)^{3/4})$, respectively, where the primes
are used for dimensional units, and the
absence thereof for dimensionless units.
The potential hereafter will be assumed
to have a customary parabolic profile~\cite{pethick2008bose,stringari} 
of the form:
\begin{equation}
    V(x)=\frac{1}{2} \Omega^2 x^2,
\end{equation}
where $\Omega$ represents the effective
strength of the
longitudinal confinement. 

In what follows we will examine
the steady state problem of Eq.~(\ref{jzeq1})
in which we will seek standing wave solutions
of the form $u(x,t)=e^{-i \mu t} q(x)$, whose
spatial profile will satisfy the steady state
equation:
\begin{eqnarray}
    \mu q= -\frac{1}{2} q_{xx}+ q^3 - \delta q^2 + V(x) q.
\label{jzeq2}
\end{eqnarray}
As is customary in the 1d setting~\cite{katsimiga2023solitary}, we will
assume (without loss of generality for our
standing solutions) that the spatial profile
is real henceforth. Recall that the homogeneous
steady states of the model are either $q=0$
or $q=(\delta \pm \sqrt{\delta^2+ 4\mu})/2$, with the
one associated with the minus sign being
modulationally unstable, while the one with
the plus sign being modulationally stable~\cite{mithun2019inter,saqlain2023dragging}.

\begin{figure}[h!]
         \includegraphics[width=0.45\textwidth]{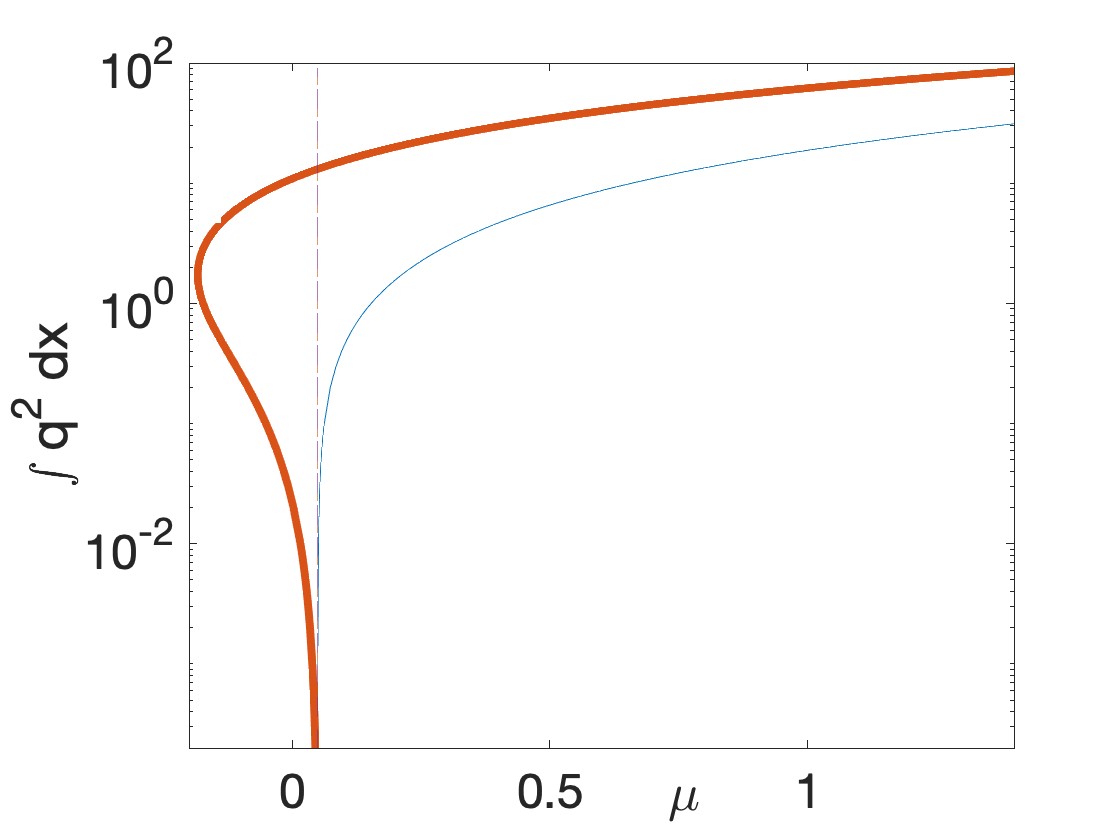}
        \includegraphics[width=0.45\textwidth]{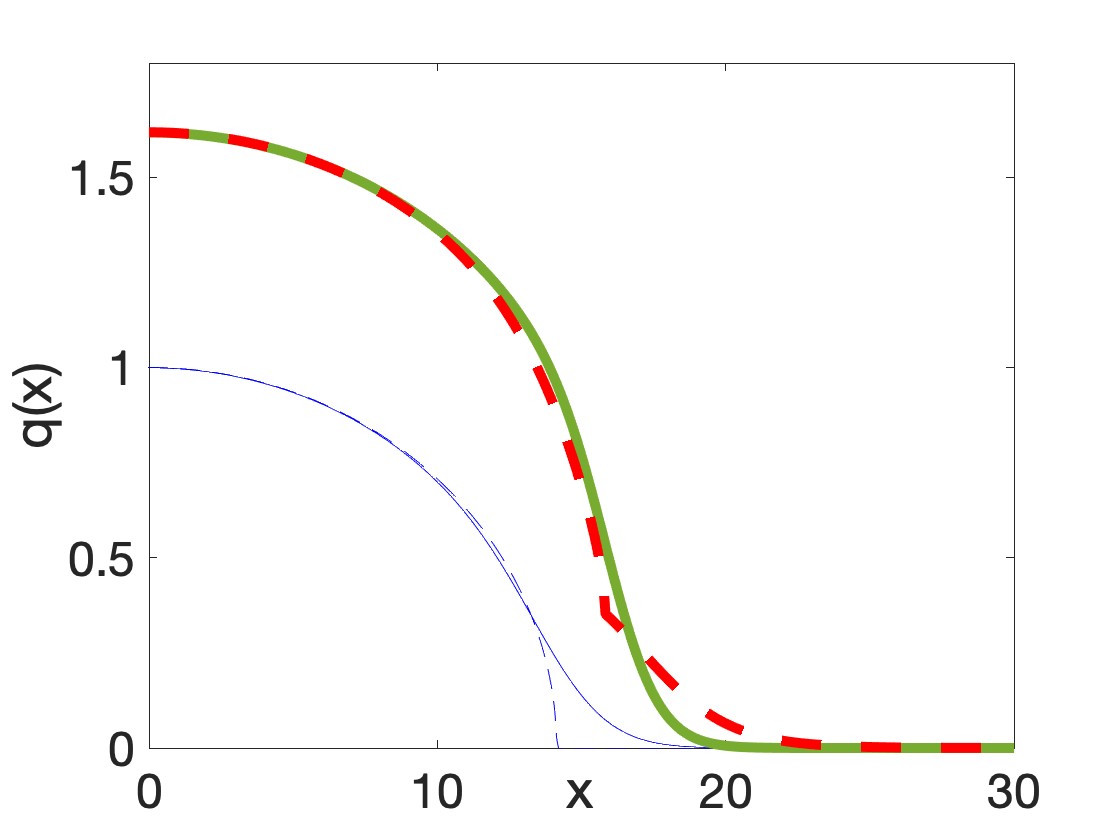}
\caption{(Top Panel) Typical example of
the ground state branch of solutions ``mass'' 
(rescaled atom number)
vs. the
chemical potential $\mu$
for a parabolic trap of strength $\Omega=0.1$
for the case of the cubic nonlinearity problem
(standard GPE, thin blue line) vs. the quadratic-cubic
nonlinearity of the present work (extended
GPE, related to the droplet problem, red
thick line). 
The vertical line denotes the linear (harmonic
oscillator) limit.
(Bottom panel) Prototypical
example of the numerical solution (solid
line) for the standard GPE (thin blue)
vs. extended GPE (thick green). The analytical
approximations of the Thomas-Fermi
inverted parabolic profile for the former
(dashed blue) and of the present work
(dashed red) are also given for comparison. Here,
the chemical potential is chosen as $\mu=1$.}.
\label{jz_fig1}
\end{figure}

Accordingly, the TF approximation in the presence
of the trap replaces $\mu$ with $\mu-V(x)$,
which can immediately be seen to be problematic
when the quantity under the radical 
$\delta^2 + 4 (\mu-V(x))=0$, at which point the density 
is finite ($q(x)=\delta/2$) and no continuous 
approximation leading to an asymptotically 
vanishing wavefunction can be constructed
(contrary to the standard cubic GPE case). 
It is this conundrum that we wish to resolve
through our analysis, providing an explicit
spatial expression of increasing accuracy 
as $\mu$ increases for the spatially confined
droplet profile. The relevant branch of 
solutions and the approach of its ``mass''
(the scaled atom number) to the linear
limit of the harmonic oscillator is
shown in Fig.~\ref{jz_fig1}. The latter
limit of asymptotically vanishing density
pertains to the ground state of
the harmonic oscillator (HO) with $\mu\rightarrow \Omega/2$; recall that, more generally, the linear
eigenstates of the HO have energies 
$\mu=(n+1/2) \Omega$,
with the ground state pertaining to $n=0$.
It can be seen that the cubic GPE problem
monotonically tends to this limit, as is
expected from its defocusing nonlinearity,
while the competing nonlinearity of the
quantum droplet problem is manifested in 
the non-monotonic approach to the relevant
limit; see also~\cite{PhysRevResearch.5.023175,katsimiga2023solitary}.
In the figure, we also show a prototypical example
of the TF limit for $\mu=1 \gg \Omega$.
Once again, for comparison the case of the 
standard GPE is shown (in blue thin
solid line), together with the (less refined
yet straightforwardly) analytically tractable
inverted parabola TF approximation (blue
thin dashed line). Here, we also include
for the same chemical potential the 
numerically exact solution of the droplet
problem, obtained via a fixed point iteration,
compared with our analytically derived 
approximate profile (both in thick lines,
the former in solid green, while the latter in
dashed red). This clearly manifests the accuracy of our analytical approximation into which we now
delve.

Changing variable
\begin{equation}
\label{E:16}
z = \frac{\sqrt 2 \, \Omega \, x}{\sqrt{\delta^2+4\mu}} \,, \qquad
w = \frac{2q - \delta}{\sqrt{\delta^2+4\mu}}
\end{equation}
gives the semiclassical form:
\begin{equation}
\label{E:01}
0 = -\epsilon^2 w_{zz} + (w^2-f^2)(w+\sigma) \,,
\end{equation}
where $f(z)^2=1-z^2$ and 
\begin{equation}
\label{E:17}
\epsilon = \frac{2\Omega}{\delta^2+4\mu} \,, \qquad \sigma = \frac{\delta}{\sqrt{\delta^2+4\mu}} \,.
\end{equation}
For $|z|\leq 1$, we take $f(z) = \sqrt{1-z^2}$, while
outside this region $f(z)=0$.

For fixed $0<\sigma<3$, we determine the form of an even solution to \eqref{E:01} asymptotically for $0< \epsilon \ll 1$.  The form of the nonlinearity in \eqref{E:01} suggests that, in regions where $w$ is slowly oscillating, if $|z|\leq 1$, we should have either $w\approx f$, $w\approx -f$, or $w\approx -\sigma$.  Based upon the assumption that there exists an even solution $w$ such that $w\approx f$ for $|z|\ll 1$ and $w \to -\sigma$ as $|z|\to \infty$, we can calculate its expected functional form.  The results are as follows, with the supporting calculations given in \S\ref{S:calc}.

\begin{figure}
\includegraphics[scale=0.6]{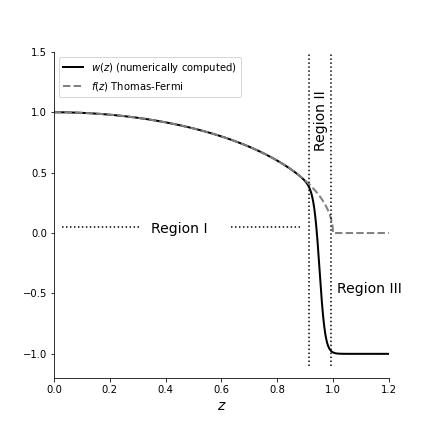}
\caption{\label{F:regions}The different regions of analysis, shown in the case $\epsilon=0.01$ and $\sigma=1$.  In this case, $z_*=0.9531$ by numerical computation.}
\end{figure}

Let $0<z_*<1$ be defined as the value of $z$ at which
\begin{equation}
\label{E:12}
w(z_*) = \frac{f(z_*)-\sigma}{2}
\end{equation}
The analysis divides into three regions (see Fig \ref{F:regions})
\begin{itemize}
\item Region I, $0\leq z \leq z_* - O(\epsilon)$, where $w\approx f$; 
\item Region II, $z_*-O(\epsilon) \leq z \leq z_*+O(\epsilon)$, where $w$ rapidly transitions from $w\approx f$ to $w\approx -\sigma$;
\item Region III, $z \geq z_*+O(\epsilon)$, where $w\approx -\sigma$.  
\end{itemize}

The asymptotic description of $z_*$ and $w$, derived in \S\ref{S:calc}, that is uniformly valid in Regions I, II and III, is:
\begin{equation}
\label{E:02}
z_* = z_0 + \frac{3\sqrt 2}{4\sigma} \epsilon + O(\epsilon^2)
\end{equation}
where $z_0 = \sqrt{1-\frac{\sigma^2}{9}}$ (so that $f_0=f(z_0) = \sigma/3$) and
\begin{equation}
    \label{E:03}
w(z) = \frac{f(z)-\sigma}{2} - \frac{f(z)+\sigma}{2} \tanh\left( \frac{2\sigma}{3\sqrt 2} \frac{z-z_*}{\epsilon}\right)
\end{equation}

As stated above, the result is derived for fixed $0<\sigma<3$ as an asymptotic expansion for $\epsilon \searrow 0$.  To determine a practical range of applicability, note that Region II should be situated strictly inside $|z|\leq 1$.  Taking $z_*$ as given by \eqref{E:02}, this converts to the condition
$$\epsilon < \frac{4\sigma}{3\sqrt 2}\left( 1- \sqrt{1-\frac{\sigma^2}{9}}\right)$$

When \eqref{E:02}, \eqref{E:03} are converted using \eqref{E:16}, \eqref{E:17}, the result takes the form
\begin{equation}
\label{E:18}
x_* = \frac{1}{\sqrt 2 \Omega}  \sqrt{\frac{8\delta^2}{9} + 4\mu} \; + \; \frac12 + O(\Omega)
\end{equation}
and
\begin{equation}
\label{E:19}
u(x) = \frac14( \delta + \sqrt{\delta^2+4\mu - 2\Omega^2 x^2})\left(1- \tanh\big(\frac{ \delta(x-x_*)}{3}\big)\right)
\end{equation}
In this form, the result can be interpreted as an asymptotic expansion for $\mu, \delta = O(1)$ and $0<\Omega \ll 1$.    The effective range of applicability converts to 
\begin{equation}
\label{E:20}\Omega < \sqrt 2\left( \sqrt{\delta^2+4\mu} - \sqrt{ \frac{8\delta^2}{9} + 4\mu} \right) \,.
\end{equation}
The upper bound \eqref{E:20} allows for a comparison to the case $\delta=0$ studied by \cite{Gallo2009OnTT}.  For $\mu=O(1)$, as $\delta \searrow 0$, \eqref{E:20} becomes $\Omega < \frac{2\delta^2}{9\sqrt{2\mu}}$, showing that $\Omega$ is pinched to zero in this limit.  Indeed, $x_*$ is approaching the boundary of the TF layer, and the refined Painlev\'e II asymptotics of \cite{Gallo2009OnTT} are needed for a more accurate description.

\section{Detailed Analysis\label{S:calc}}

To justify \eqref{E:02}, \eqref{E:03}, define $\phi$ via the equation
\begin{equation}
\label{E:04}
w = \frac{f-\sigma}{2} - \frac{f+\sigma}{2}\phi
\end{equation}
Let $y = (z-z_0)/\epsilon$.   In this reference frame, Region I corresponds to $y \ll -1$, Region II lies in $-1\lesssim y \lesssim 1$, and Region III corresponds to $y \gg 1$.  Comparing \eqref{E:12} and \eqref{E:04}, we see that $z_*$ is characterized as the $z$-value at which $\phi(z_*)=0$.

In Region I, $w\approx f$, so that in view of \eqref{E:04}, we fix the $y\to -\infty$ boundary conditions as
\begin{equation}
\label{E:07}
\phi \to -1 \,, \quad \phi_y \to 0 \, \quad \text{as} \quad y\to -\infty
\end{equation}
In Region III, $w\approx -\sigma$, so in view of \eqref{E:04}, we set the $y\to +\infty$ boundary conditions as
\begin{equation}
\label{E:08}
\phi \to 1 \,, \quad \phi_y \to 0 \, \quad \text{as} \quad y \to +\infty
\end{equation}
By assuming that the boundary conditions \eqref{E:07}, \eqref{E:08} hold up to second order in $\epsilon$, we will be able to obtain a consistent expansion, as follows.

Plugging \eqref{E:04} into \eqref{E:01} and changing variable $y \mapsto \xi$, with
\begin{equation}
\label{E:11}
\frac{d\xi}{dy} = \frac{f+\sigma}{2\sqrt 2} \, \qquad \xi=0 \leftrightarrow y=0
\end{equation}
leads to the transformed equation
\begin{equation}
\label{E:06}
\begin{aligned}
&\phi_{\xi\xi} + 2(1-\phi^2)(\phi-h) \\
&\qquad =- \epsilon \frac{12 f_z}{\sqrt 2(f+\sigma)^2} \phi_\xi + \epsilon^2 \frac{8 f_{zz}}{(f+\sigma)^3}(1-\phi)   
\end{aligned}
\end{equation}
where $h= (3f-\sigma)/(f+\sigma)$.  Multiplying \eqref{E:06} by $\phi_\xi$ and integrating gives
\begin{equation}
\label{E:10}
H(\phi,\phi_\xi)\Big]^{\xi=+\infty}_{\xi=-\infty} = \int_{-\infty}^{+\infty} R(\xi) \, d\xi
\end{equation}
where
\begin{align}
\notag
&H(\phi,\phi_\xi) = \tfrac12 \phi_\xi^2 + \phi^2 - \tfrac12 \phi^4 \\
\label{E:05}
&R = 2h(1-\phi^2) \phi_\xi  - \epsilon \frac{12 f_z}{\sqrt 2 (f+\sigma)^2} \phi_\xi^2
\end{align}
For the boundary values \eqref{E:07}, \eqref{E:08}, the left side of \eqref{E:10} is zero. 

The equation  \eqref{E:06} is still an exact representation of \eqref{E:01}, but at this point we initiate an asymptotic expansion.   In order that the right side of \eqref{E:10} vanish at zero order in $\epsilon$, we need $h=O(\epsilon)$.  Setting $h=0$ and dropping the right side of \eqref{E:06}, it becomes $0 = \phi_{\xi\xi} + 2(1-\phi^2)\phi$, which has the solution 
\begin{equation}
\label{E:09}
\phi(\xi) = \tanh(\xi-\xi_1) \,,
\end{equation}
where $\xi_1$ is an undetermined shift.  Setting $h=0$ is equivalent to setting $f=\sigma/3$.  Let
$$z_0 = \sqrt{1-\frac{\sigma^2}{9}} \quad \iff 
 \quad f(z_0)=\frac{\sigma}{3} \,.$$
With $z_0$ as a point of reference, the asymptotic form of \eqref{E:11} is:
\begin{equation}
\label{E:13}
y = \frac{2\sqrt 2}{f_0+\sigma}\xi + O(\epsilon)\xi^2 =\frac{3\sqrt{2}}{2\sigma} \xi + O(\epsilon) \xi^2
\end{equation}
The expansion of $h$ is
$$h = - \epsilon \frac{81 \sqrt 2 z_0}{8\sigma^3} \xi + O(\epsilon^2) \xi^2 $$
This and using and \eqref{E:09} as an approximation for $\phi$ in \eqref{E:05} provides an expansion for $R$:
$$R = - \epsilon \frac{81 z_0}{2\sqrt 2 \sigma^3} (\xi-\tfrac12) \sech^4(\xi-\xi_1) +O(\epsilon^2)$$
Substituting this into the right side of \eqref{E:10} (with left side $=0$) implies that $\xi_1=\frac12+O(\epsilon)$ in order that the right side of \eqref{E:10} vanish at first order in $\epsilon$.

Substituting \eqref{E:13} into $y=(z-z_0)/\epsilon$, we obtain
\begin{equation}
\label{E:14}
z = z_0 + \epsilon \frac{3\sqrt 2}{2\sigma} \xi + O(\epsilon^2) \xi^2
\end{equation}
Recall that $z_*$ is characterized as the $z$-value at which $\phi(z_*)=0$.  By \eqref{E:09}, this corresponds to $\xi=\xi_1$ and thus
\begin{equation}
\label{E:15}
z_* = z_0 + \epsilon \frac{3\sqrt 2}{2\sigma} \xi_1 + O(\epsilon^2)
\end{equation}
which gives \eqref{E:02}.  From \eqref{E:09}, we can replace $\phi$ with $\tanh$ in \eqref{E:04} and also use the difference of \eqref{E:14} and \eqref{E:15} to reexpress $\xi-\xi_1$, in order to obtain  \eqref{E:03}. 

\section{An Extension: Dark Solitons\label{S:ext}}

\begin{figure}[h!]
         \includegraphics[width=0.45\textwidth]{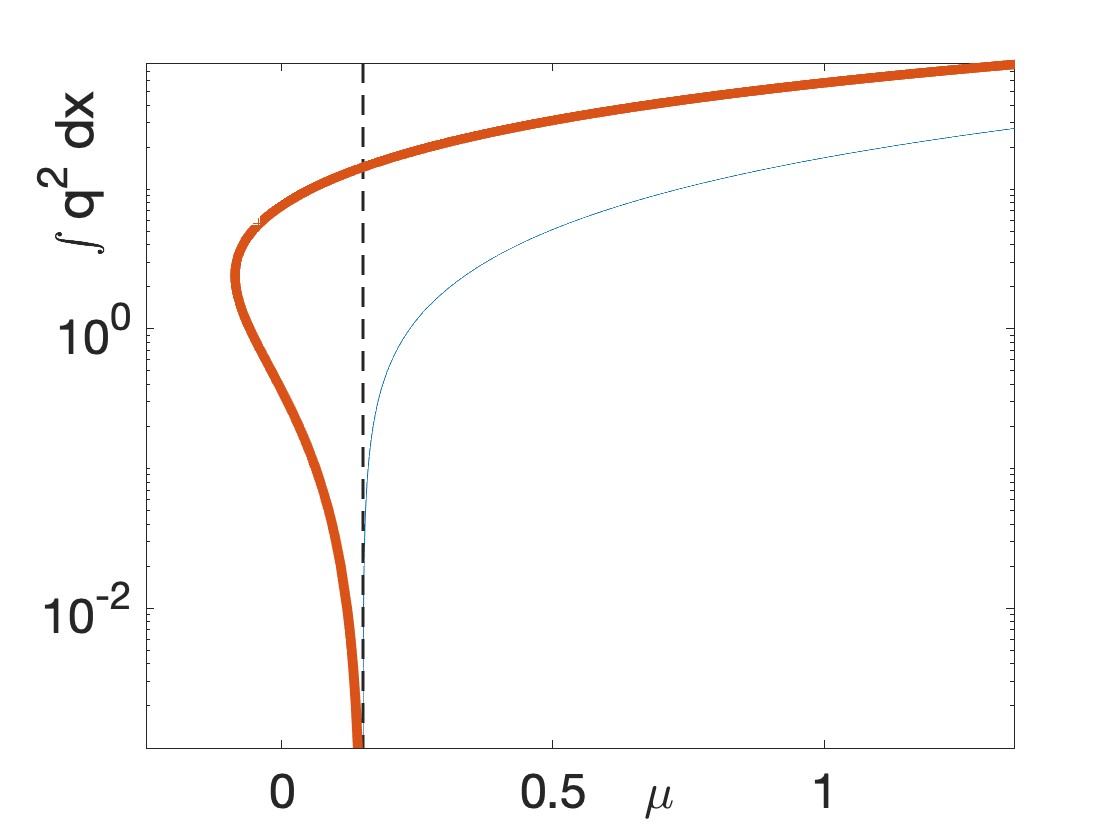}
        \includegraphics[width=0.45\textwidth]{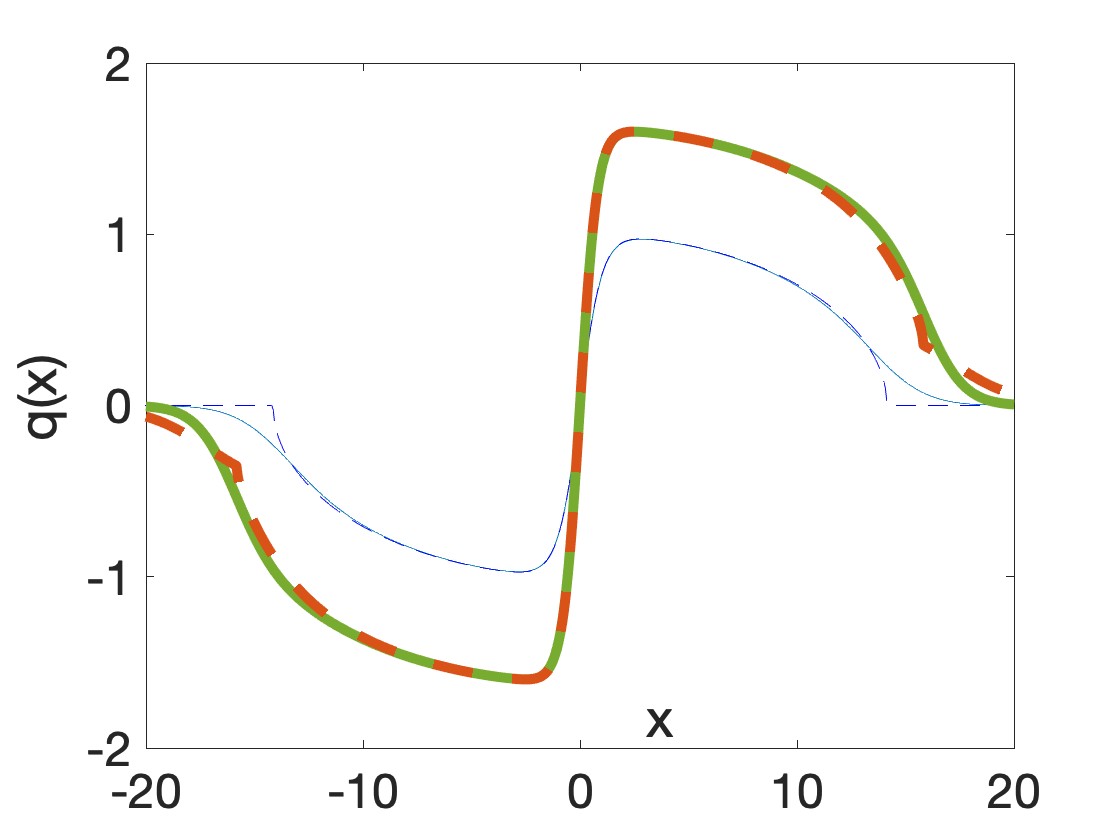}
\caption{(Top panel) Similar to 
Figure~\ref{jz_fig1}, but now for the
prototypical dark soliton structure.
Once again, the trap strength is 
$\Omega=0.1$, hence the linear limit of the
(harmonic oscillator) first excited state
lies at $\mu=3\Omega/2=0.15$. The thin blue
line of the GPE is compared to the red thick
line of the droplet case dark soliton branch
continuation. (Bottom panel)
Comparison of the respective dark solitons.
The thin blue lines reflect the GPE result:
here, the solid line shows the numerical dark soliton,
while the dashed one represents the Thomas-Fermi
approximation multiplied by the $\tanh$ 
soliton profile).
The respective droplet model states, also
for $\mu=1$ (as in the GPE) are shown by thick
lines, with the analytical (dashed line) state arising from
the multiplication of Eq.~(\ref{E:19}) by
the analytical solitary wave of Ref.~\cite{katsimiga2023solitary}.}
\label{jz_fig2}
\end{figure}

Having constructed a uniform approximation
to the ground state of the quantum droplet
model in the large $\mu$ limit, we now 
turn to an interesting extension of the
relevant waveform. In particular, it is well-known
that for the standard GPE model, the
excited states in the form of dark solitons
can be well-approximated in the large density
(large chemical potential) limit by the ground
TF state multiplied by the dark soliton
of the homogeneous equation~\cite{Coles_2010}.
A relevant example for chemical potential
$\mu=1$ and $\Omega=0.1$ is shown in the
bottom panel of Figure~\ref{jz_fig2}.
The central portion of the corresponding
one-soliton stationary, anti-symmetric 
excited state wavefunction (the so-called
black soliton~\cite{frantzeskakis2010dark})
is very well approximated by $\tanh(x)$, while
the remaining waveform, aside from the boundary
layer discussed in section I, is well approximated
by the TF profile.

It is then natural to expect that a similar
strategy can be used to {\it analytically}
approximate, in a uniform way, black solitons
in the quantum droplet model, which have been
the subject of intense recent research efforts~\cite{PhysRevResearch.5.023175,katsimiga2023solitary,condmat8030067}. In particular, 
we leverage the analytical approximation 
of Eq.~(\ref{E:19}) for the ground state
and multiply the relevant spatial profile by the
exact analytical dark soliton of the
homogeneous quantum droplet setting, as derived
in~\cite{katsimiga2023solitary}. 
Indeed, we use a similar notation as that
work labeling $q_+=(\delta + \sqrt{\delta^2+ 4 \mu})/2$ and expressing the black soliton as:
\begin{eqnarray}
\label{DSsol}
u_{\rm \,dark}(x)=q_++\frac{-{\cal B}(\mu) + \sqrt{{\cal B}^2(\mu)-4 {\cal A}(\mu) {\cal C}(\mu)}}{2 {\cal A}(\mu)},\qquad
\end{eqnarray}
in which expression the symbols 
${\cal A}$, ${\cal B}$ and ${\cal C}$ are given by:
${\cal A}(\mu)=B^2-4 A \tanh^2(\sqrt{A} (x))$, 
${\cal B}(\mu)=4 A B \sech^2(\sqrt{A} (x))$, and
${\cal C}(\mu)=4 A^2 \sech^2(\sqrt{A} (x))$, 
with $A=4 \mu + (1+\sqrt{1+4 \mu})$ and
$B=2 (\frac{1}{3}+\sqrt{1+4 \mu})$.
It is important to recall that 
the expression of Eq.~(\ref{DSsol}) 
can be used only for $x$ such that $u_{\rm dark}>0$,
while the profile is supposed to be 
anti-symmetric around the point of zero crossing.
Notice that also the point of zero crossing
in the homogeneous model can be shifted at will
due to the translational invariance of the underlying setting. 
By centering the relevant dark soliton
around the center of the trap, as is expected
for the stationary trapped black soliton state,
and multiplying it by Eq.~(\ref{E:19}),
we observe in the bottom panel
of Fig.~\ref{jz_fig2} that we get a very
accurate approximation (thick dashed line)
to the full numerical result (thick solid line).
This happens for large chemical potentials
($\mu \gg \Omega$) for the branch of dark
solitary wave solutions that is shown in the
top panel of the figure, once again compared
between the GPE and the EGPE models, in order
to observe the impact of the attractive 
(beyond-mean-field) nonlinear term in the latter.
This, in turn, can be the basis for
analyses similar to those of~\cite{Coles_2010}
that may enable the systematic characterization
of excited (multiple dark solitary wave) states stability
and dynamics. We comment on this 
possibility further in the
next section.

\section{Conclusions and Future Challenges}

In this work we have analyzed the ground
state of the extended Gross-Pitaevskii equation
in the presence of a parabolic confinement
in the regime of the so-called Thomas-Fermi
limit, i.e., for the case of large densities/chemical
potentials. We combined a separation of the
spatial domain into different regions
(the central region, the rapid transition
---interface--- region and the asymptotic state
region) along with an analysis of each one 
through suitable rescalings and 
asymptotic methods in order to extract
a uniformly valid asymptotic formula
that we tested again in direct numerical
computations via fixed point iteration methods
to provide an increasingly accurate description
of the quantum droplet in the large density
regime. 

Naturally, this development paves the way for
a number of possible considerations for the future.
On the one hand, this naturally poses
the question of whether approximate 
excitation frequencies for the quantum
droplet can be extracted in this limit,
by analogy of what has been done for the case
of the standard GPE; see, e.g., the discussion
of~\cite{kevdep}. On the other hand, the
availability of such an analytical ``ansatz''
for a TF solution may offer the backdrop for
the consideration of the asymptotic form
of higher excited states, such as 
multiple dark solitons,
in analogy with earlier works that were
able to derive effective particle equations
for such coherent structures~\cite{Coles_2010}. 
At the same time, such analysis can provide
a starting point for the consideration of
higher dimensional analogues of the model
and the asymptotic analysis of both 
droplet, but also importantly vortical
patterns therein~\cite{saqlain2023dragging,yongyao}.
Such studies are currently in progress and 
will be presented in future publications.

\acknowledgements
This material is based upon work supported by the U.S.\ National Science 
Foundation under the awards PHY-2110030 and DMS-2204702 (PGK), and  DMS-2055072 (JH).
PGK gratefully acknowledges numerous 
insightful discussions with S.I. Mistakidis,
G.C. Katsimiga, R. Carretero-Gonz{\'a}lez, S. Chandramouli
B.A. Malomed, G.N. Koutsokostas  and D.J. Frantzeskakis 
on the subject of quantum droplets.

\bibliographystyle{apsrev4-1}
\bibliography{references}

\end{document}